\documentclass[twocolumn,prl,showpacs]{revtex4}

\usepackage{graphicx}
\usepackage{dcolumn}
\usepackage{float}
\usepackage{amsmath}

\newcommand{\comment}[1]{}

\begin{document}

\title{Exchange Boson Dynamics in Cuprates: Optical Conductivity of
HgBa$_{2}$CuO$_{4+\delta}$}

\author{J. Yang$^{1}$}
\altaffiliation[Current address: ]{School of Materials Science and Engineering, Tianjin University, Tianjin 300072, P. R. China. }
\author{J. Hwang$^{1}$}
\altaffiliation[Current address: ]{Department of Physics, University of Florida,
Gainesville, Florida 32611, USA.}
\author{E. Schachinger$^{2}$}
\author{J. P. Carbotte$^{1,5}$}
\author{R.P.S.M. Lobo$^{3}$}
\author{D. Colson$^{4}$}
\author{A. Forget$^{4}$}
\author{T. Timusk$^{1,5}$}
\email{timusk@mcmaster.ca}
\affiliation{$^{1}$Department of Physics and Astronomy,
McMaster University, Hamilton, ON L8S 4M1, Canada\\
$^{2}$Institute of Theoretical and Computational
Physics, Graz University of Technology, A-8010 Graz, Austria\\
$^{3}$Laboratoire Photons et Mati\`{e}re, CNRS UPR5, LPS-ESPCI, 10 rue Vauquelin, 75231
Paris Cedex 5, France\\
$^{4}$CEA, IRAMIS, SPEC, 91191 Gif sur Yvette, France\\
$^{5}$The Canadian Institute of Advanced Research, Toronto, Ontario M5G 1Z8, Canada.}

\date{\today}


%
%

\begin{abstract}
The electron-boson spectral density function $I^{2}\chi(\Omega)$ responsible for carrier
scattering of the high temperature superconductor HgBa$_{2}$CuO$_{4+\delta}$ ($T_{c}$ =
90 K) is calculated from new data on the optical scattering rate.  A maximum entropy
technique is used. Published data on 
HgBa$_{2}$Ca$_{2}$Cu$_{3}$O$_{8+\delta}$ ($T_{c}$ = 130 K) are also inverted and these
new results are put in the context of other known cases. All spectra (with two notable
exceptions) show a peak at an energy ($\Omega_{r}$) proportional to the superconducting transition temperature $\Omega_{r}\approx$
6.3$k_{B}T_{c}$.  This charge channel relationship follows closely the magnetic resonance seen by polarized neutron scattering, $\Omega_r^{neutron}\approx$ 5.4$k_{B}T_{c}$. The
amplitudes of both peaks decrease strongly with increasing temperature. In
some cases, the peak at $\Omega_r$ is weak and the spectrum can have
additional maxima and a background extending up to several hundred meV.

\end{abstract}


\maketitle

Superconductivity arises when electronic quasiparticles bind together into Cooper pairs
which condense into a macroscopic quantum state. The interaction which drives the pairing
could be charge or spin polarizations, but in conventional metals, it is the polarization
of the lattice of ions which provides the necessary attraction. There is also a smaller
Coulomb repulsion. In the phonon case Eliashberg theory accounts for all the experimental
data with its central ingredient, the electron-phonon spectral density
$\alpha^{2}F(\Omega)$ specifying the boson exchange pairing
interaction~\cite{carbotte90}.  Optical spectroscopy is a powerful tool that can be used
to extract bosonic spectral density
with high degree of accuracy in a variety of
systems where other techniques such as tunneling and angle resolved photoemission (ARPES) are difficult to apply.  It
has been used in conventional metals ~\cite{farnworth74,marsiglio98} and in the high
$T_{c}$
oxides~\cite{carbotte99,schachinger00,schachinger06,hwang06,hwang07,hwang08,schachinger}
but its interpretation in the cuprates remains controversial. As emphasized recently by
Anderson~\cite{anderson07}, the pairing interaction could involve high energy virtual
transitions across the Mott gap with energy set by the Hubbard $U$ which is a large
energy and the effective interaction would be instantaneous on the time scale of
interest. On the other hand, recent numerical work~\cite{maier08,kyung08} based on the
\textit{t-J} model, has shown that the main contribution to the pairing glue is provided
by the spin fluctuations with characteristic energies of at most a few hundred meV.
Optical experiments provide a direct probe of this energy region.

The mercury versions of the cuprates, HgBa$_{2}$CuO$_{4+\delta}$ (Hg1201) and
HgBa$_{2}$Ca$_{2}$Cu$_{3}$O$_{8+\delta}$ (Hg1223) provide a unique opportunity to test
these ideas.  The one-copper layer system Hg1201 shares a conventional transition
temperature around 90 K with widely studied systems  YBCO and Bi2212, whereas the three-layer
compound Hg1223 has a $T_c= 130$ K, nearly 40 \% higher.  These dramatically different
$T_c$'s lead, as we will show, to very different bosonic spectra and place severe
constraints on models of superconductivity in the cuprates.

In this letter, we present new data on the optical constants for
HgBa$_{2}$CuO$_{4+\delta}$ (Hg1201). The bosonic spectral density $I^{2}\chi(\Omega)$,
recovered by maximum entropy inversion, is found to have a remarkable resemblance to
previous results for optimally doped Bi2212~\cite{hwang07}. For further comparison, using
the same methods, we also invert published~\cite{mcguire00} optical constants in the
three-layer Hg1223. The high-quality Hg1201 single crystal was grown by a flux growth
technique. Our sample was a millimeter sized platelet with a well oriented
\textit{ab}-plane. It is slightly underdoped with a $T_c = 91$ K. The real and imaginary parts of the optical conductivity
$\sigma(T,\omega)\equiv\sigma_1(T,\omega)+i\sigma_2(T,\omega)$ follow from the
reflectivity. In analyzing optical data, we use a memory function or optical self energy
$\Sigma^{op}(T,\omega)$ instead of working with
$\sigma(T,\omega)$\cite{puchkov96,mori00}. By definition,
$\sigma(T,\omega)=(i\omega_{p}^2/4\pi)/[\omega-2\Sigma^{op}(T,\omega)]$ where
$\omega_{p}$ is the plasma frequency. The optical self energy defined this way plays a
role analogous in optics (which involves a two particle process) to the quasiparticle
self energy $\Sigma^{qp}(T,\omega)$ in ARPES~\cite{johnson01,hwang07b}. The optical scattering rate is
$1/\tau^{op}(T,\omega)=-2\Sigma_{2}^{op}(T,\omega)$ and the optical effective mass
$m^{op}(T,\omega)/m$ (with $m$ the bare electron mass) is given by
$\omega[m^{op}(T,\omega)/m-1]\equiv-2\Sigma_{1}^{op}(T,\omega)$.

%
%
\begin{figure}[t!]
  \includegraphics[width=9cm]{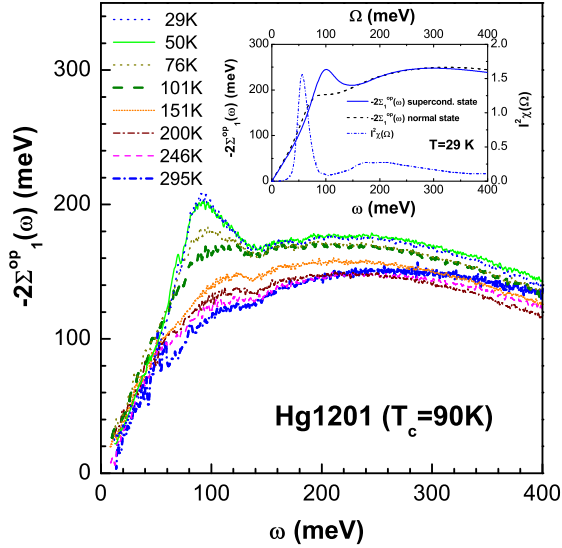}
\caption{(color online). Experimental results for the real part of the optical self
energy $-2\Sigma_{1}^{op}(T,\omega)$ as a function of photon energy $\omega$ for eight
temperatures. Inset, theoretical results based on numerical solutions of the generalized
Eliashberg equations. (Solid blue superconducting and dashed black normal state.) The
electron-boson exchange spectral density used is shown as the dashed-dotted blue curve.
The superconducting gap value is $\Delta$ = 22.4 meV.} \label{fig1}
\end{figure}

Our results for the real part of the optical self energy $-2\Sigma_{1}^{op}(T,\omega)$,
which provides an easy comparison with ARPES, often presented in terms of
$2\Sigma_{1}^{qp}(T,\omega)$, are shown in Fig. \ref{fig1} as a function of $\omega$ for
eight temperatures. The lowest three temperatures are in the superconducting state. The
most prominent feature of the curves is the peak around 100 meV seen at $T$ = 29 K, which
progressively disappears as the temperature is increased. To better see how the bosonic
spectrum generates the self energy we show in the inset of the figure a model calculation
based on numerical solutions of the $d$-wave Eliashberg equations. The input
electron-boson spectral density $I^{2}\chi(\Omega)$, shown as the dashed-dotted curve,
consists of a large narrow peak (right hand scale) centered at 56 meV followed by a dip
and a long background extending beyond 400 meV. The dashed curve, {\it with the same
bosonic spectrum}, is the normal state result for $-2\Sigma_{1}^{op}(T,\omega)$ which is
to be compared with the solid curve in the superconducting state. The superconducting gap
obtained was 22.4 meV giving a gap to $T_{c}$ ratio, $2\Delta/k_{B}T_{c}$ = 5.8. Note
that, as theory predicts~\cite{carbotte05}, the dashed curve in the normal state shows no
visible structure at $\omega=\Omega_{r}= $ 56 meV. Instead there should be zero slope at
$\omega=\sqrt{2}\Omega_{r}$ for an Einstein spectrum. It is clear that boson structure is
hard to see in normal state. However in the superconducting states the quasi-particle
electronic density of states acquires energy dependence and this helps reveal the
underlying boson structure as seen in the solid blue curve. The peak at $\sim$100 meV is
neither at $\Omega_r$ nor $\sqrt{2}\Omega_{r}$ but is shifted upwards by the opening of
the gap $\Delta$~\cite{carbotte05} but as we see, the position of the peak in
$I^{2}\chi(\Omega)$ cannot be read off the curve without the knowledge of the value of
the superconducting gap.

%
%
\begin{figure}[t!]
  \includegraphics[width=3.5in]{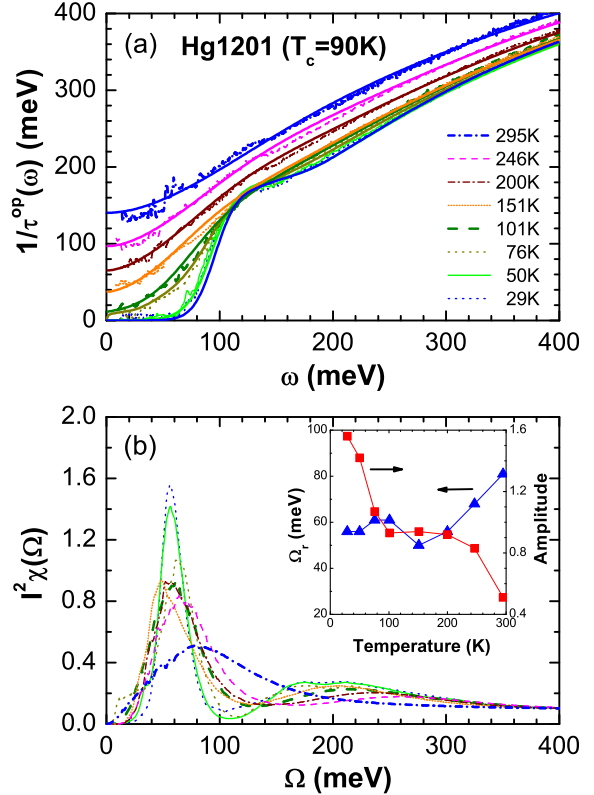}
\caption{(color online). Top frame, the optical scattering rate $1/\tau^{op}(T,\omega)$
for Hg1201 v.s. $\omega$ for 8 temperatures (light curves). The wider curves are our
maximum entropy reconstructions. Bottom frame, the electron-boson spectral function
$I^2\chi(\Omega)$ v.s. $\Omega$. The inset gives the peak position (blue triangles) left
scales as a function of temperature and the red squares give the corresponding peak
amplitude.} \label{fig2}
\end{figure}

In Fig. \ref{fig2}(b) we show results for the spectral density $I^{2}\chi(\Omega)$ of
maximum entropy inversions augmented with a least squares improvement based on the full
\textit{d}-wave Eliashberg equations~\cite{schachinger06}. Further applications are found
in Refs.~\cite{hwang07,hwang08,schachinger}. The input to the inversion is the optical
scattering rate $1/\tau(\omega)={\omega_p^2\over {4\pi}}{\cal R}{e}({1} /
\sigma(\omega))$.
These are shown in Fig. \ref{fig2}(a) where the results of our inversions (heavy lines)
are compared with the original data (light lines). In all cases, the fit is very good.
Recently, van Heumen \textit{et al.}~\cite{vanheumen,vanheumen07} have also presented
optical data for Hg1201 above $T_c$ which they analyze in terms of a spectral density
represented by a set of histograms.
 While they obtain fits which are of equal quality to
ours and have a background extending to $> 400\,$meV as we have,
they find that the height of the peak at 56 meV does not change contrary to our
findings for $T$ = 246 and 295 K. This has been taken as evidence
for coupling of the charge carriers to phonons \cite{lee2006}.
We can also get fits where the peak height does not change but only if we use a {\it biased} maximum
entropy inversion with the default model set to the previous lower temperature solution
instead of being set to the constant of the {\it unbiased} inversion.

In the inset to the lower frame of Fig. \ref{fig2}, we show the frequency (left scale,
triangles) of the prominent peak in $I^{2}\chi(\Omega)$ as a function of temperature. As also 
noted by van Heumen \textit{et al}.~\cite{vanheumen}, $\Omega_{r}$ is fairly constant at
$\sim$ 56 meV but in our analysis, this frequency clearly increases for $T$ above 200 K.
More importantly, the amplitude of the peak shows strong temperature dependence in the
superconducting state and also above 200 K. Further, the width of the peak increases with
increasing $T$. As noted for Bi2212~\cite{hwang07}, the shift in spectral weight into the
peak at 60 meV can be interpreted to proceed through a transfer of spectral weight from
high to low frequencies as the temperatures is lowered.  We note  here that in contrast a bosonic function from the electron phonon interaction would not have these properties: its amplitude, width and center frequency would all be temperature independent and would not vary from one cuprate to another (as shown in Fig. 4 below). 

%
%
\begin{figure}[t!]
  \includegraphics[width=3.5in]{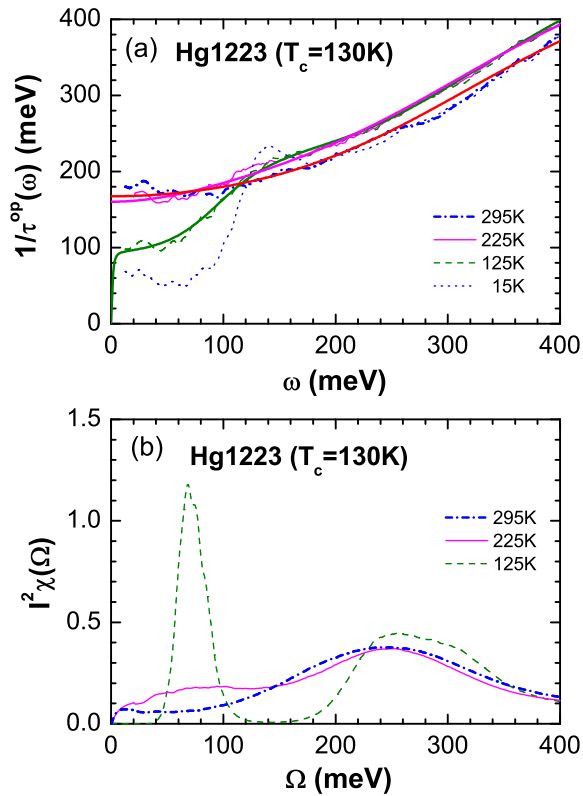}
\caption{(color online). Same as Fig. \ref{fig2} but for Hg1223 ($T_c$=130 K).}
\label{fig3}
\end{figure}

In Fig.~\ref{fig3} we display similar results for Hg1223 with $T_c=130\,$K. In Fig.~%
\ref{fig3}(a) we reproduce the optical scattering rate at four temperatures (with $T=15$
K and $125$ K in the superconducting state) from the work of McGuire {\textit{et
al.}}~\cite{mcguire00}. Our results for $I^2\chi(\Omega)$ are presented in Fig.
\ref{fig3}(b) and the quality of the data reconstruction is demonstrated by the heavy
lines in Fig. \ref{fig3}(a) to be compared with the corresponding light line (data). We
find significant residual (static impurity $1/\tau_{\rm imp} \simeq 95\,$meV) scattering
rate in contrast to Hg1201 which is in the clean limit ($1/\tau_{\rm imp}=0$).
The superconducting state data (blue, dotted curve)
for $1/\tau^{\rm op}(\omega)$ at $T=15\,$K shows a peak around $140\,$meV which is the
indication of a gap in the charge carrier density of states (DOS). Currently there is no
known kernel which allows maximum entropy inversion of such data
and so we do not show results in this instance. On the other hand, in
the $T=125\,$K data there is no signature of such a DOS-gap.
Results are shown in Fig.~\ref{fig3} b). The prominent peak
at $\Omega_r$ $\cong$ 72 meV seen in the curve at $T$ = 125 K is missing at higher
$T$. In contrast to the Hg1201 case no reasonable alternate fits can be found for $T$ =
225 K and 295 K which show a significant peak amplitude at 72 meV.

%
%
\begin{figure}[t!]
  \includegraphics[width=7cm]{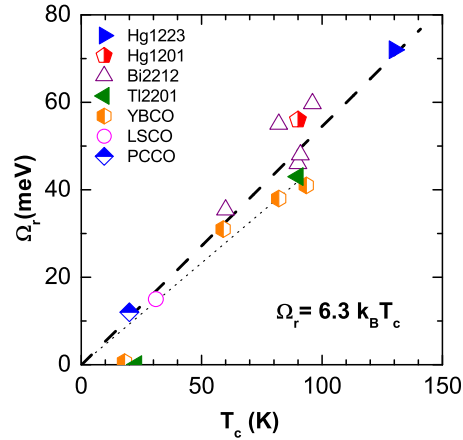}
\caption{(color online). The optical resonance frequency $\Omega_r$ as a function of
$T_c$. Bi2212~\cite{hwang07,schachinger06}, Tl2201~\cite{schachinger00},
YBCO~\cite{schachinger00,schachinger06,hwang06}, LSCO~\cite{hwang08}, and
PCCO~\cite{schachinger}.} \label{fig4}
\end{figure}

In Fig. \ref{fig4}, we place our results for $\Omega_r$, the frequency of the peak in $I^{2}\chi(\Omega)$, in the context of other such results by plotting $\Omega_r$ as a function of the
superconducting $T_c$ for a number of cuprates. In all cases, $I^{2}\chi(\Omega)$ has been extracted from the
optical data but not always using a maximum entropy technique. Some are fits to assumed
forms including a broad background introduced in reference~\cite{millis90} to model
antiferromagnetic fluctuations, augmented with a resonance peak. Both methods give very
much the same results as documented in reference~\cite{schachinger00}. The heavy long
dashed line is a least square fit to all the optical data and gives $\Omega_{r}\approx$
6.3$k_{B}T_{c}$. This is close to, but not quite, the position of the spin-one neutron resonance  obtained by He \textit{et al.}~\cite{he01,he02} where $\Omega_{r}^{neutron} \approx$ 5.4$k_{B}T_{c}$ represented in Fig. \ref{fig4} as the
dotted line. 

Several comments should be made about such a comparison between charge
excitations and the magnetic susceptibility.  First, the the neutron resonance plotted in  Fig. \ref{fig4} refers to the sharp peak that appears at $q=(\pi,\pi)$ whereas the bosonic function that governs the optical response is the $q$ averaged local susceptibility. This arises because the Fermi-surface to Fermi-surface electron scattering involves momentum transfers to boson excitations that
span all momenta in the Brillouin zone, some involving Umklapp processes. Where magnetic neutron scattering data for the $q$-averaged susceptibility are available such as the Ortho II YBCO ~\cite{stock05} or optimally doped LSCO~\cite{vignolle07} the agreement between the neutron data and the optical data is excellent: not only are the peaks in the response at the same frequencies but also the temperature dependence of the amplitude of the peaks are in agreement~\cite{hwang07,hwang08}.    Secondly, it should be noted that in some cuprates, the resonance described involves only a very small fraction of the
total weight seen in the local spin susceptibility as in LaSrCuO~\cite{hwang08} and
PrCeCuO~\cite{schachinger}. In YBCO$_{6.5}$ it is estimated to be 3\%~\cite{stock05}
and in some cases there is no resonance~\cite{schachinger00,stock06}, but it
is always the local, Brillouin zone averaged spin susceptibility which controls superconductivity. 
Finally, the two points at $\Omega_{r}$ = 0 (not used in the fit to the data in Fig.
\ref{fig4}) are for YBCO$_{6.35}$~\cite{stock06} and overdoped
Tl2201~\cite{schachinger00}. In both cases, no optical resonance could be identified.  The
resonance may enhance but is not essential for superconductivity in the cuprates and the scaling of the position of the peak with $T_c$ shown in Fig. \ref{fig4}
must be the result and not the cause of the rearrangement of the electronic DOS in the superconducting
state as suggested by several theorists~\cite{abanov99,prelovsek06}.  We also note here that recent dynamical mean field calculations of the one-band Hubbard model yield bosonic spectral functions very similar to what is shown in Figs. 2 and 3.~\cite{maier08,kyung08}

In summary we find that in Hg1201 and Hg1223 optical resonances are found in maximum entropy inversions of the optical scattering, at 56 and 72 meV, respectively. However, when the temperature
is increased towards 300 K, the spectral weight under this resonance moves to higher energy and broadens significantly, in contrast to
the findings of van Heumen \textit{et al.}~\cite{vanheumen07}. The optical resonance scales with $T_c$ over a broad set of materials with $\Omega_r\approx 6.3k_BT_c$ which is remarkably
close to the energy of the spin one resonance seen in polarized
neutron scattering, namely $\Omega^{neutron} = 5.4k_BT_c$ leaving no doubt
that the charge carriers are coupled to spin fluctuations, while there is no evidence for an important phonon contribution. 

This work has been supported by the Natural Science and Engineering Research
Council of Canada and the Canadian Institute for Advanced Research.
RPSML acknowledges support from the ANR grant BLAN07-1-183876 GAPSUPRA.

Note added in proof.ÑWe have learned of a neutron 
scattering study 4 
by Yu et al. [31], where a magnetic reso- 
nance in optimally doped Hg1201 is reported at 56 meV, 
exactly the same energy as the peak we found here. Recent 
Raman data 5 
Þnd a superconducting gap 2?? very close to 
the values found in our calculations [32].

{\it Note added in proof}. -We have learned of a neutron scattering study by Yu {\it et al}~\cite{yu08}, where a magnetic resonance in optimally doped Hg1201 is reported at 56 meV, exactly the same energy as the peak we found here. Recent Raman data finds a superconducting gap $2 \Delta$ very close to the values found in our calculations.\cite{guyard08} 

%
%


\begin{thebibliography}{99}

\bibitem{carbotte90} J.P. Carbotte, Rev. Mod. Phys \textbf{62}, 1027 (1990).

\bibitem{farnworth74} B. Farnworth and T. Timusk, \prb \textbf{10}, 2799 (1974).

\bibitem{marsiglio98} F. Marsiglio, {\it et al.}, Phys. Lett. A \textbf{245}, 172 (1998).


\bibitem{carbotte99} J.P. Carbotte, {\it et al.}, Nature
(London) \textbf{401}, 354 (1999).

\bibitem{schachinger00} E. Schachinger and J.P. Carbotte, \prb \textbf{62},
9054 (2000).

\bibitem{schachinger06} E. Schachinger, {\it et al.}, \prb
\textbf{73}, 184507 (2006).

\bibitem{hwang06} J. Hwang, {\it et al.}, \prb \textbf{73}, 014508 (2006).

\bibitem{hwang07} J. Hwang, {\it et al.}, \prb \textbf{75}, 144508 (2007).

\bibitem{hwang08} J. Hwang, {\it et al.}, \prl \textbf{100}, 137005 (2008).

\bibitem{schachinger} E. Schachinger, {\it et al.}, \prb {\bf 78,} 134522 2008.

\bibitem{anderson07} P. W. Anderson, Science {\bf 316}, 1705 (2007).

\bibitem{maier08} T.A. Maier, {\it et al.} Phys. Rev. Letters {\bf 100}, 237001 (2008).

\bibitem{kyung08} B. Kyung {\it et al.}, arXiv:0812.1228.


\bibitem{mcguire00} J.J. McGuire, {\it et al.}, \prb \textbf{62}, 8711 (2000).

\bibitem{puchkov96} A.V. Puchkov, {\it et al.}, J. Phys. Condens.
Matter \textbf{8}, 10049 (1996).

\bibitem{mori00} T. Mori, {\it et al.}, \prb \textbf{77}, 174515 (2008).

\bibitem{johnson01} P.D. Johnson, {\it et al.}, \prl \textbf{87}, 177007 (2001).

\bibitem{hwang07b} J. Hwang, {\it et al.}, \prl \textbf{98}, 207002 (2007).

\bibitem{carbotte05} J.P. Carbotte, {\it et al.}, \prb \textbf{71}, 054506 (2005).

\bibitem{vanheumen} E. van Heumen, {\it et al.}, arXiv:0807.1730.

\bibitem{vanheumen07} E. van Heumen, {\it et al.}, \prb \textbf{75}, 054522 (2007).
  
\bibitem{lee2006}J. Lee {\it et al.} Nature (London) 442, 546 (2006).

\bibitem{millis90} A.J. Millis, {\it et al.}, \prb \textbf{42}, 167 (1990).

\bibitem{he01} H. He, {\it et al.}, \prl \textbf{86}, 1610 (2001).

\bibitem{he02} H. He, {\it et al.}, Science \textbf{295}, 1045 (2002).

\bibitem{stock05} C. Stock, {\it et al.}, \prb \textbf{71}, 024522 (2005).

\bibitem{vignolle07} B. Vignolle, {\it et al.}, Nature Phys. \textbf{3}, 163 (2007).


\bibitem{stock06} C. Stock, {\it et al.}, \prb \textbf{73}, 100504(R) (2006).

\bibitem{abanov99} Ar. Abanov and A.V. Chubukov, \prl \textbf{83,} 1652 (1999).
\comment{A Relation between the resonance peak and ARPES Data in the Cuprates}

\bibitem{prelovsek06} P. Prelov\v sek and I. Sega, \prb \textbf{74,}  214501 (2006).

\bibitem{yu08} G. Yu, {\it et al.}, arXiv:0810.5759.

\bibitem{guyard08} W. Guyard, {\it et al.} Phys. Rev. Lett. {\bf 101,} 097003 (2008).



\end{thebibliography}
\end{document}